\documentclass[
aps,
twocolumn,
superscriptaddress,
amsmath,amssymb,
pre
]{revtex4-1}
\usepackage{graphicx}
\usepackage{dcolumn}
\usepackage{bm}

\begin{document}
\title{Brownian motion at short time scales}

\author{Tongcang Li}
  \email{Now at the University of California, Berkeley}
 \affiliation{Center for Nonlinear Dynamics and Department of Physics,
The University of Texas at Austin, Austin, TX 78712, USA}
\author{Mark G. Raizen}
  \email{Corresponding author;\quad E-mail: raizen@physics.utexas.edu}
\affiliation{Center for Nonlinear Dynamics and Department of Physics,
The University of Texas at Austin, Austin, TX 78712, USA}

\date{\today}
\begin{abstract}
 Brownian motion has played important roles in many different fields of science since its origin was first explained by Albert Einstein in 1905. Einstein's theory of Brownian motion, however, is only applicable at long time scales.  At short time scales, Brownian motion of a suspended particle is not completely random, due to the inertia of the particle and the surrounding fluid. Moreover, the thermal force exerted on a particle suspended in a liquid is not a white noise, but is colored. Recent experimental developments in optical trapping and detection have made this new regime of Brownian motion accessible. This review summarizes related theories and recent experiments on Brownian motion at short time scales, with a focus on the measurement of the instantaneous velocity of a Brownian particle in a gas and the observation of the transition from ballistic to diffusive Brownian motion in a liquid.
\end{abstract}

\maketitle

\section{Introduction}

Brownian motion is the apparently perpetual and random movement of particles suspended in a fluid (liquid or  gas), which
  was first observed systematically  by Robert Brown  in 1827 \cite{brown1828}. When Brown used a simple microscope  to study the action of particles from  pollen immersed in water \cite{brown1828},  he ``observed many of them very evidently in motion''. The size of those particles was about 5 $\mu$m.  He also observed the same kind of motion with powders of many other materials, such as wood and nickel, suspended in water.

As first  explained by  Einstein in 1905  \cite{einstein1905}, the Brownian motion of a suspended particle is a  consequence of the thermal motion of surrounding fluid molecules. Einstein's theory of Brownian motion predicts that
 \begin{equation}
\label{MSDeistein}
\langle{[\Delta x(t)]^2}\rangle \equiv \langle{(x(t)-(x(0))^2}\rangle = 2 D  t ,
\end{equation}
where $\langle{[\Delta x(t)]^2}\rangle$ is the mean-square displacement (MSD)   of a free Brownian particle in one dimension during time  $t$
, and $D$ is the diffusion constant. The diffusion constant can be calculated by $D = k_B T/\gamma$, where $k_B$ is the Boltzmann constant, $T$ is the temperature, and $\gamma = 6 \pi \eta R$ is the Stokes friction coefficient for a sphere with radius $R$. Here $\eta$ is the viscosity of the fluid.

M. von Smoluchowski also derived the expression of MSD independently in 1906 \cite{smoluchowski1906}, with a result that differed from Eq. (\ref{MSDeistein}) by a factor of about 2. In 1908, Paul Langevin introduced a stochastic force  and derived Eq. (\ref{MSDeistein}) from Newton's second law \cite{langevin1908,lemons1997}. Langevin's approach is much more intuitive than Einstein's approach, and the resulting ``Langevin equation'' has found broad applications in stochastic physics \cite{Coffey1996}. Experimental confirmation of Eq. (\ref{MSDeistein}) was provided by the brilliant experiments of Jean Perrin \cite{perrin1910}, recognized by the Nobel Prize in Physics in 1926. Theodor Svedberg also verified the Einstein-Smoluchowski theory of Brownian motion and won the Nobel Prize in Chemistry in 1926 for related work on colloidal systems \cite{nobel1966}.


Persistence and randomness are generally accepted as two key characteristics of  Brownian motion. The trajectories of  Brownian particles are classic examples of fractals \cite{Vicsek1992}. They are commonly assumed to be continuous  everywhere but not differentiable anywhere \cite{Nelson1966}. Since its trajectory is not differentiable, the velocity of a Brownian particle is undefined. According to Eq. (\ref{MSDeistein}),  the  mean velocity measured over an interval of time $t$ is   $\bar{v} \equiv \sqrt{\langle{[\Delta x(t)]^2}\rangle}/ t =  \sqrt{2D}/\sqrt{t}$. This  diverges as $t$ approaches 0, and therefore does not represent the real velocity of the particle \cite{einstein1907,einstein1956}.

\begin{figure}[b!]
\setlength{\unitlength}{1cm}
\begin{picture}(8,4.5)
\put(0,0){\includegraphics[totalheight=4.3cm]{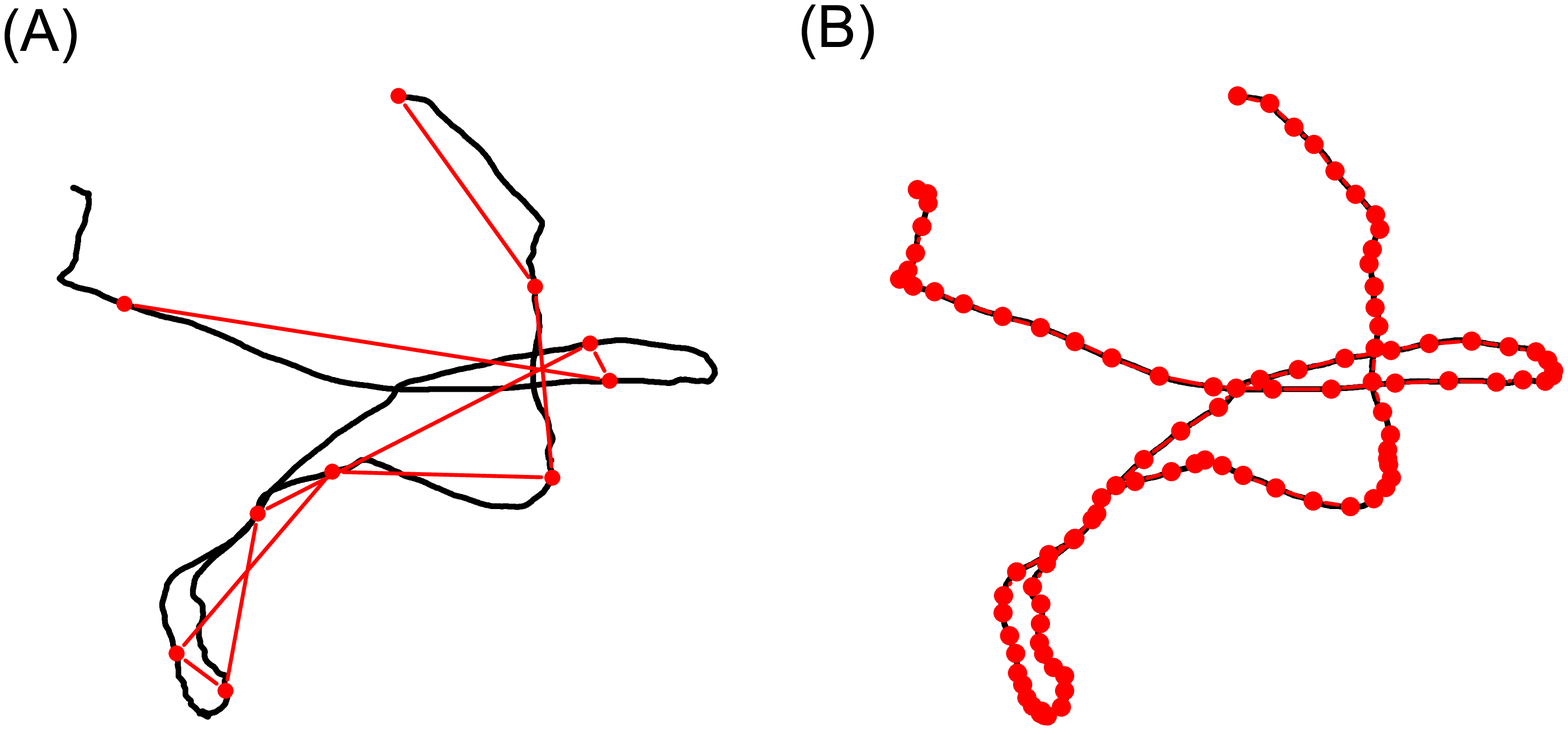}}
\put(2.5,0.5){\large $\vec{v}\neq\frac{\Delta \vec{x}}{\Delta t}$}
\put(6.5,0.5){\large $\vec{v}\doteq\frac{\Delta \vec{x}}{\Delta t}$}
\end{picture}
\caption[A 2D trajectory of a  Brownian particle]{\label{fig-1-Brownian}  A 2D trajectory of a  Brownian particle. The black curve is assumed to be a true trajectory of the particle. Red dots are measured positions, and red curves are measured trajectories. The sampling rate of (B) is 10 times of that of (A). }
\end{figure}

At short time scales ($t \ll \tau_p$, where $\tau_p = M/\gamma$ is the momentum relaxation time of a particle with mass $M$),   the dynamics of a Brownian particle is expected to be dominated by its inertia and its trajectory cannot be self-similar. This is termed ``ballistic Brownian motion'' to be distinguished from the common ``diffusive Brownian motion''.
Fig. \ref{fig-1-Brownian} shows a 2D trajectory of a  Brownian particle. The black curve is assumed to be a true trajectory of the particle. Red dots are measured positions. In Fig. \ref{fig-1-Brownian}A, the sampling rate is too small to measure the velocity of the Brownian particle. The measured trajectory (red curve) is completely different from the true trajectory, and appears chaotic. It is impossible to obtain the velocity of the particle from the measured trajectory in Fig. \ref{fig-1-Brownian}A. In Fig. \ref{fig-1-Brownian}B, the sampling rate is much larger.  Now the measured trajectory is very close to the true trajectory of the particle. If the measured displacement of the particle is $\Delta \vec{x} (t)$ during time $\Delta t$, then the velocity of the particle is approximately $\vec{v} = \Delta \vec{x} (t) / \Delta t$.

In 1900, F. M. Exner made the first quantitative study of Brownian motion by measuring the velocity of  Brownian  particles suspended in water \cite{exner1900,kerker1974}. He found that the measured velocity decreased with increasing particle size and increased with increasing water temperature. However, his measured velocities   were almost 1000-fold smaller than those predicted by the energy  equipartition theorem \cite{exner1900}.
The reason of this discrepancy was not understood until A. Einstein developed his kinetic theory about Brownian motion \cite{einstein1905}.

In 1907, Einstein published a paper entitled ``Theoretical observations on the Brownian motion'' in which he considered the instantaneous velocity of a Brownian particle \cite{einstein1907,einstein1956}.  Einstein showed that by measuring this quantity, one could prove that ``the kinetic energy of the motion of the centre of gravity of a particle is independent of the size and nature of the particle and independent of the nature of its environment''.  This is one of the basic tenets of statistical mechanics, known as the equipartition theorem.  However, Einstein concluded that due to the very rapid randomization of the motion, the  instantaneous velocity of a Brownian particle would be impossible to measure in practice \cite{einstein1907,einstein1956}:

\renewcommand{\thefootnote}{\roman{footnote}}

``We must conclude that the velocity and direction of motion of the particle will be already very greatly altered in the extraordinary short time $\theta$, and, indeed, in a totally irregular manner. It is therefore impossible -- at least for ultramicroscopic particles -- to ascertain $\sqrt{\overline{v^2}}$ by observation."

Einstein's conclusion was unchallenged for more than 100 years because of the exclusive difficulty of such a measurement. For a 1 $\mu$m diameter silica (SiO{\scriptsize 2}) sphere in water at room temperature, the momentum relaxation time $\tau_p$ is about 0.1 $\mu$s and the root mean square (rms) velocity $v_{rms}=\sqrt{k_B T/M}$ is about 2 mm/s in one dimension. To measure the instantaneous velocity with 10\% uncertainty, one would require 2 pm spatial resolution and 10~ns temporal resolution, which is a very difficult task. Due to the lower viscosity of gas as compared to liquid, the momentum relaxation time $\tau_p$ of a particle in air is much larger. This lowers the requirements of both temporal and spatial resolution.


Nondiffusive Brownian motion of colloidal suspensions with high concentrations at short time scales have been studied by  measuring the autocorrelation
functions of multiply scattered, transmitted light \cite{weitz1989,zhu1992,kao1993}.
Recent developments in optical tweezers and  detection systems with unprecedented resolution now prove to be an indispensable tool for studying the Brownian motion of a single particle at short time scales \cite{lukic2005,leonardo2007,chavez2008,li2010,li2011b,burnham2010,huang2011,franosch2011,Pusey2011}.
For example,
 we have  observed the ballistic Brownian motion and measured the instantaneous velocity of a Brownian particle for the first time with an optically trapped bead in air \cite{li2010}.
 Huang \emph{et al}  have observed the transition from ballistic Brownian motion to diffusive Brownian motion in a liquid \cite{huang2011}.
 Franosch \emph{et al} have observed resonances arising from hydrodynamic memory in Brownian motion in a liquid and the long-sought colored  spectrum of the thermal force \cite{franosch2011,bergsorensen2005}.

Besides Brownian motion at short time scales, theoretical and experimental studies of anisotropic Brownian motion and Brownian motion in nonequilibrium systems are currently pursued by many groups. For example, several groups reported anisotropic Brownian motion
of  particles near interfaces \cite{holmqvist2006,tinoco2007,jeney2008,michailidou2009,wang2009}, and Brownian motion of
anisotropic particles such as ellipsoids \cite{han2006,Mukhija2007}, nanotubes \cite{tsyboulski2008,fakhri2010} and helical bacteria \cite{Butenko2012}.  Brownian motion in nonequilibrium systems is of particular interest because it is directly related to the transport of molecules and cells in biological systems.  Important examples include Brownian motors \cite{astumian1997,hanggi2009}, active Brownian motion of self-propelled particles \cite{peruani2007,lindner2008,golestanian2009,jiang2010,mino2011,romanczuk2011,selmeczi2007},  hot Brownian motion \cite{rings2010}, and Brownian motion in shear flows \cite{ziehl2009}. Recent theoretical studies also found that the inertias of particles and  surrounding fluids  can significantly affect the Brownian motion in nonequilibrium systems \cite{Yuge2010,Ghosh2012a,Ghosh2012b,Celani2012,Chakraborty2012,Hottovy2012}.

In this review, section 2 introduces the theories of Brownian motion of particles in a gas, and the recent measurement of the instantaneous velocity of a Brownian particle in air. Section 3 introduces the theories of Brownian motion of particles in a liquid at short time scales,  the experimental observation of the colored thermal force, and the transition from ballistic to diffusive Brownian motion in a liquid.
Section 4 discusses the effects of detection noise on the measurement of different quantities of Brownian motion.
Finally, in section 5, we discuss future experiments on Brownian motion at short time scales.

\section{Brownian motion in a gas}

In this section, we assume that the density of a gas is much smaller than the density of the suspended Brownian particles. So the inertia effects of the gas can be neglected.

\subsection{Theory}

The mean free path of  molecules in air at 1 atmosphere at room temperature is about 68 nm \cite{jennings1988}. The collision rate between a 1-$\mu$m-diameter microsphere suspended in air and surrounding air molecules is  about $10^{16}$ Hz at ambient conditions. The observed Brownian motion  is an averaged effect of these ultrafast collisions. Because of the huge difference between the mass of a microsphere and that of an air molecule, the motion of a microsphere can only be changed significantly by a large number of collisions. This is reflected in the fact that $\tau_p = 6 \,\mu$s for  a 1-$\mu$m-diameter silica  microsphere in air at ambient conditions.
The dynamics of a Brownian particle  at time scales much longer than that of individual collisions  can be described by a Langevin equation  \cite{langevin1908,lemons1997}.  Here we introduce the theory of Brownian motion in a gas following the classic work of Uhlenbeck and Ornstein \cite{uhlenbeck1930}.

\subsubsection{A free particle in a gas}
 The dynamics of a Brownian particle with mass $M$ in a gas can be described by a Langevin equation \cite{langevin1908,lemons1997,uhlenbeck1930}:
 \begin{equation}
\label{eq4-2}
M \frac{d^2 x}{dt^2} + \gamma \frac{dx}{dt} = F_{therm}(t),
\end{equation}
 where
  \begin{equation}
\label{eq4-3}
F_{therm}(t)=(2 k_B T \gamma)^{1/2} \zeta(t)
\end{equation}
 is the Brownian stochastic force.
$\zeta(t)$ is a normalized white-noise process. Hence for all $t$ and $t'$,
 \begin{equation}
\label{eq4-4}
\langle\zeta(t)\rangle =0 \text{ , and } \langle \zeta(t)\zeta(t')\rangle =\delta(t-t').
\end{equation}

The MSD for a Brownian particle at thermal equilibrium with the air is \cite{uhlenbeck1930}:
 \begin{equation}
\label{eq4-12}
\langle{[\Delta x(t)]^2}\rangle  = \frac{2 k_B T}{M \Gamma^2_0} (\Gamma_0 \, t-1+e^{- \Gamma_0 \, t}),
\end{equation}
where $\Gamma_0 = \gamma /M $ is the damping coefficient. We have $\tau_p = 1/\Gamma_0$.
At long time scales ($t \gg \tau_p$), the MSD is the same as that predicted by Einstein's theory (Eq. (\ref{MSDeistein})).
At very short time scales, the MSD is
 \begin{equation}
\label{eq4-14}
\langle{[\Delta x(t)]^2}\rangle = \frac{k_B T}{M} t^2 \quad \text{ for }\quad t \ll \tau_p .
\end{equation}
The velocity autocorrelation function is \cite{uhlenbeck1930}:
 \begin{equation}
\label{eq4-9}
\langle v(t)v(0)\rangle = \frac{k_B T}{M} e^{- \Gamma_0 t} .
\end{equation}

Although these equations are initially derived for an ensemble of  particles, the ergodic theorem dictates that they are also valid for measurements of a single particle taken over a long time.

The damping coefficient $\Gamma_0$ in a gas can be calculated by kinetic theory. Assuming the reflection of gas molecules from the surface of a microsphere is diffusive, and the molecules thermalize  with the surface during collisions, we obtain \cite{li2011a,beresnev1990}
\begin{equation}
\label{eq3} \Gamma_0 = \frac{6 \pi \eta R}{M}\frac{0.619}{0.619+\texttt{Kn}} (1+c_K),
\end{equation}
where $\eta$ is the viscosity coefficient of the gas, $R$ is the radius of the microsphere, and $\texttt{Kn} = s/R$ is the Knudsen number. Here $s$ is the mean free path of the gas molecules. $c_K=(0.31 \texttt{Kn})/(0.785+1.152 \texttt{Kn}+\texttt{Kn}^2)$ is a small positive function of $\texttt{Kn}$ \cite{beresnev1990}. At high pressures where $\texttt{Kn} \ll 1$, the  damping coefficient is $\Gamma_0 = 6 \pi \eta R /M$, which is the same as the prediction of Stokes' law.

At very short time scales, the motion is ballistic and
its instantaneous velocity can be measured as $v = \Delta x(t)/ t$, when $t \ll \tau_p$  \cite{uhlenbeck1930}. The ballistic Brownian motion is different from a simple ballistic motion.
For a simple ballistic motion with velocity $u$, we have $\Delta x(t)=u t$ and $[\Delta x(t)]^2=u^2 t^2$. The velocity $u$ can be any value and usually has no relation with the temperature of the environment. For the ballistic Brownian motion, the amplitude of the velocity is determined by the temperature of the environment.
The 1D Maxwell-Boltzmann distribution of the velocity of a particle in thermal equilibrium is
 \begin{equation}
\label{eq4-14-1}
f_v(v_i)= \sqrt{\frac{M}{2 \pi k_B T}} \exp{\left(-\frac{M v^2_i}{2 k_B T}\right)},
\end{equation}
where $v_i$  is the velocity of the particle along direction $i$, which can be any direction.

\subsubsection{An optically trapped microsphere in a gas}

For small displacements, the effect of optical tweezers on the microsphere's motion can be approximated by that of a harmonic potential. The Brownian motion of a particle in a harmonic trap has been studied by Uhlenbeck and Ornstein \cite{uhlenbeck1930}, and Wang and Uhlenbeck \cite{wang1945}.
The  equation of the Brownian motion of a microsphere in a harmonic trap is:
 \begin{equation}
\label{eq4-15}
 \frac{d^2 x}{dt^2} + \Gamma_0 \frac{dx}{dt} + \Omega^2 x= \Lambda \zeta(t),
\end{equation}
where $\Omega=\sqrt{\kappa/m}$ is the natural angular frequency of the trapped microsphere when there is no damping, and $\Lambda=(2 k_B T \Gamma_0 /M)^{1/2}$. The cyclic frequency of the damped oscillator is $\omega_1 = \sqrt{\Omega^2-\Gamma^2_0/4}$. The  system is underdamped when $\omega_1$ is real ($\Omega>\Gamma/2$), critically damped when $\omega_1=0$, and overdamped when $\omega_1$ is imaginary ($\Omega<\Gamma/2$).

The MSD of a Brownian particle in an underdamped harmonic trap in air is \cite{wang1945}:
\begin{equation}
\label{eq4-17} \langle{[\Delta x(t)]^2}\rangle = \frac{2 k_B T}{M \Omega^2} \left[ 1-e^{-t /2\tau_p} \left( \cos \omega_1 t + \frac{\sin \omega_1  t}{2 \omega_1 \tau_p}  \right) \right].
\end{equation}
The position autocorrelation function is related to the MSD by:
\begin{equation}
\label{eq4-18}
\langle{[\Delta x(t)]^2}\rangle = 2\langle{x^2}\rangle -2\langle{x(t)x(0)}\rangle,
\end{equation}
where $\langle{x^2}\rangle = k_B T/(M\Omega^2)$. The rms amplitude is $x_{rms}=\sqrt{k_B T/(M\Omega^2)}$.
The normalized position autocorrelation function (PACF) of the particle is \cite{wang1945}:
\begin{equation}
\label{eq4-19}  \frac{\langle x(t)x(0)\rangle}{\langle x^2\rangle}= e^{- t /2\tau_p} \left( \cos \omega_1  t + \frac{\sin \omega_1 t}{2 \omega_1 \tau_p}  \right).
\end{equation}
The normalized velocity autocorrelation function (VACF) of the particle is \cite{wang1945}:
\begin{equation}
\label{eq4-20}  \frac{\langle v(t)v(0)\rangle}{\langle v^2\rangle}= e^{- t /2\tau_p} \left( \cos \omega_1  t - \frac{\sin \omega_1 t}{2 \omega_1 \tau_p}  \right).
\end{equation}
Both the position autocorrelation function and the velocity autocorrelation function oscillate for an underdamped system.

Similar to the optical spectrum of an atom, the power spectrum of the Brownian motion of a trapped microsphere contains a lot of information about the system. The power spectral density (PSD) of a variable is  the squared modulus of its Fourier transform \cite{wang1945,bergsorensen2004,norrelykke2011}.
The expected values of the PSD of an optically trapped microsphere in air is
\begin{equation}
\label{eq4-29}
S(\omega) \equiv <S^{rec}_k> = \frac{2 k_B T}{M \Omega^2} \frac{\Omega^2\Gamma_0}{(\Omega^2-\omega^2)^2+  \omega^2 \Gamma^2_0}.
\end{equation}
Eq. (\ref{eq4-29}) is valid for both underdamped \cite{li2011a} and overdamped systems.
The measured PSD of a recorded $x(t)$ is $S^{rec}_k$, which depends on a sample of the white noise $\zeta(t)$.  Thus an experimental PSD will appear noisy. Averaging many measured $S^{rec}_k$ will result in a spectrum close to the expected  spectrum $S(\omega)$. Another way to reduce the noise in a measured spectrum is ``blocking"\cite{bergsorensen2004}. A ``block" of consecutive data points ($\omega_{k_1}$,$S^{rec}_{k_1}$)... ($\omega_{k_2}$,$S^{rec}_{k_2}$) can be represented by a single new ``data point" ($\overline{\omega_k}$,$\overline{S^{rec}_k}$) which are the block averages.

\subsection{Experimental observation of the instantaneous velocity of a Brownian particle in air}
Because of the lower viscosity of air as compared to that of liquid, a particle suspended in air is an ideal system for studying the ballistic Brownian motion. The main difficulty of studying Brownian motion in air, however, is that the particle will fall under the influence of gravity.
To overcome this problem, Fedele \emph{et al} used aerosol suspensions with small particles ($\sim 0.2 \, \mu$m diameter) to achieve a long sedimentation time \cite{fedele1980}, Blum  \emph{et al} performed the experiment under microgravity conditions in the Bremen drop tower \cite{blum2006}.
Due to the lack of a detection system with a sufficient resolution, neither experiment was able to measure the instantaneous velocity of Brownian motion. We overcame this problem by using optical tweezers to simultaneously trap and monitor a silica bead in air and vacuum, allowing long-duration, ultra-high-resolution measurements of its motion. Here we review this experiment that was originally reported in Ref. \cite{li2010} with  more experimental details.

\subsubsection{A fast detection system}
In order to measure the instantaneous velocity of the Brownian motion of a trapped microsphere in air, we implemented an ultrahigh resolution detection system.

\begin{figure}[b!]
\setlength{\unitlength}{1cm}
\begin{picture}(8,6)
\put(0.3,0){\includegraphics[totalheight=6cm]{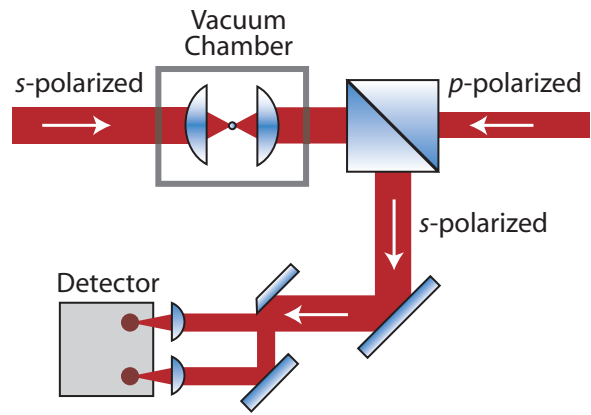}}
\end{picture}
\caption[1D split detector]{\label{fig-III-11}  This simplified schematic shows our counter-propagating dual-beam optical tweezers, and a novel detection system (Figure adapted from Ref. \cite{li2010}). The $\it s$-polarized beam  is reflected by a polarizing beam splitter cube after it passes through a trapped bead inside a vacuum chamber.  Then, for detection, it is split by a mirror with a sharp edge. The $\it p$-polarized beam passes through the cube. }
\end{figure}

We used an ultra-stable NPRO laser (Model: 126-1063-700, Lightwave Electronics (now JDSU)) to trap and monitor a silica bead  in vacuum. Its rms intensity noise is $<$ 0.05~\% over the range from 10 Hz to 2 MHz, and is shot noise limited above 10 MHz. It is a single frequency laser with a wavelength of 1064 nm and a coherence length longer than 1000 m. A detailed characterization of this type of laser can be found in Ref. \cite{kwee2008}. We  used this laser for both trapping and detection. This is achieved by using a polarizing beam splitter cube to reflect one of the trapping beams for detection (Fig. \ref{fig-III-11}).

Our lab has previously developed a fast position-sensitive laser beam detector \cite{chavez2008}. The previous detector used a  bundle of optical fibers that spatially splits the incident beam, and a fast balanced photodetector to measure the difference between the two halves of the beam.
We simplified the detection system by using a mirror with a sharp edge (BBD05-E03, Thorlabs) to replace the fiber-optic bundle for splitting the beam (Fig. \ref{fig-III-11}). The sharp edge of a mirror is much smoother than the boundary between the two halves of a fiber  bundle. So it is much simpler and has less noise  than a fiber bundle for splitting the laser beam.

We used a balanced detector (PDB120C, photodiode diameter: 0.3 mm, Thorlabs) with a bandwidth of 75 MHz for detection. The detector is sensitive to  light with wavelengths in the range of 800-1700 nm. It has a high transimpedance gain of $1.8 \times 10^5$ V/A. The detector measures the difference between the two halves of the beam, which is proportional to the particle excursion. The intensity noise of the laser is contained in both halves and is thus canceled in the measurement. This detection system enables us to monitor the real-time position of a trapped microsphere in air with {\AA}ngstrom spatial resolution and microsecond temporal resolution \cite{li2010}. We have also developed a 3D detection system that can  monitor
the 3D motion of a microsphere trapped in vacuum with a sensitivity of about
39 fm Hz$^{-1/2}$ over a wide frequency range \cite{li2011a}.

\subsubsection{Experimental results}

 A simplified scheme of our setup for measuring the instantaneous velocity of a Brownian particle in air is shown in Fig. \ref{fig-III-11}.  The trap is formed  inside a vacuum chamber by two counter-propagating laser beams focused to the same point by two identical aspheric lenses with focal length of 3.1 mm. The two 1064 nm laser beams are orthogonally polarized, and their frequencies differ by 160 MHz to avoid interference.

\begin{figure}[b!]
\setlength{\unitlength}{1cm}
\begin{picture}(8,6.8)
\put(0.0,0){\includegraphics[totalheight=6.8cm]{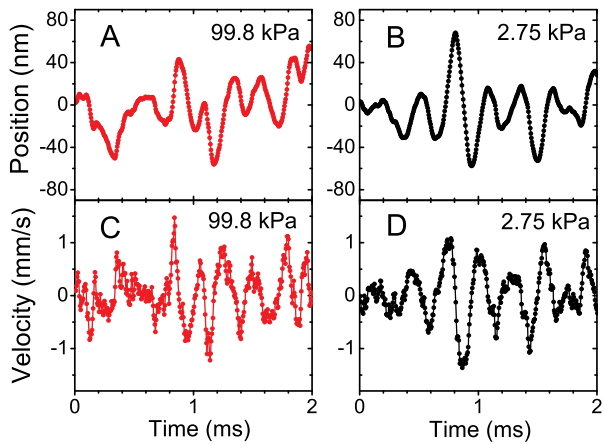}}
\end{picture}
\caption[Measured instantaneous position and velocity of a bead]{\label{measuredtrajectory}  One-dimensional trajectories of a 3 $\mu$m diameter silica bead trapped in air at 99.8 kPa ({\bf A}) and at 2.75 kPa ({\bf B}).  The instantaneous velocities of the bead corresponding to these trajectories are shown in ({\bf C}) and ({\bf D}). Figure adapted from Ref. \cite{li2010}.}
\end{figure}

 The two laser beams are aligned with the help of a pinhole aperture whose diameter is $1.0 \pm 0.5$ $\mu$m.   We intentionally make the waist of one beam larger than the other to make this alignment less critical. The measured waists of the two beams  are about 2.2  $\mu$m and 3.0  $\mu$m, respectively. Once a bead is trapped, we keep the power of one beam constant, and tune the power of the other beam to maximize the trapping frequency.

 When the bead deviates from the center of the trap, it deflects both trapping beams. We monitor the position of the bead by measuring the deflection of one of the beams, which is split by a mirror with a sharp edge.  This simple, yet novel, detection scheme has a bandwidth of 75 MHz and ultra-low noise \cite{libbrecht2004,chavez2008}.
The position signal of a trapped bead is recorded at a sampling rate of 2 MHz. Because of the detection noise, we are not able to obtain accurate instantaneous velocities of a bead at this rate. To reduce the noise, we average every 10 successive position measurements, and use these averages to calculate instantaneous velocities with time resolution of 5 $\mu$s. Although this method reduces the temporal resolution by a factor of 10, it  greatly increases the signal-to-noise ratio if  both the trapping period ($2 \pi / \omega_0$) and momentum relaxation time  are much larger than 5 $\mu$s. These conditions are satisfied here since the trapping period is about 320 $\mu$s, $\tau_p = 48$ $\mu$s at 99.8 kPa (749~torr), and $\tau_p = 147$~$\mu$s at 2.75~kPa (20.6~torr).

\begin{figure}[b!]
\setlength{\unitlength}{1cm}
\begin{picture}(8,6)
\put(0,0){\includegraphics[totalheight=6cm]{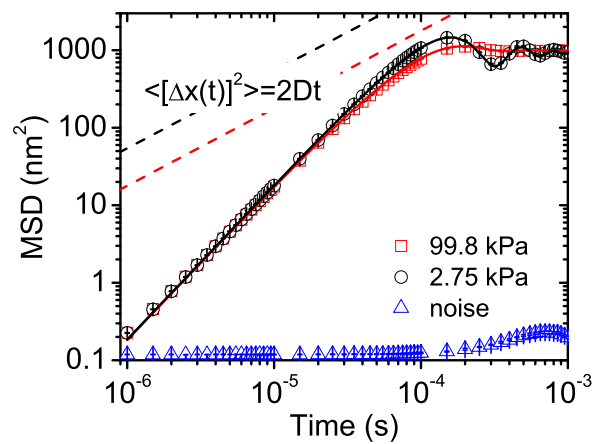}}
\end{picture}
\caption[Measured mean square displacement of a bead]{\label{MSDinair}    The mean square displacements of a 3 $\mu$m silica bead trapped in air at 99.8 kPa (red square) and 2.75 kPa (black circle).   The solid lines are the theoretical predictions of Eq. \ref{eq4-17}.  The prediction of Einstein's theory of free Brownian motion in the diffusive regime is shown in dashed lines for comparison.  Figure adapted from Ref. \cite{li2010}.}
\end{figure}

\begin{figure}[b!]
\setlength{\unitlength}{1cm}
\begin{picture}(8,6)
\put(0,0){\includegraphics[totalheight=6cm]{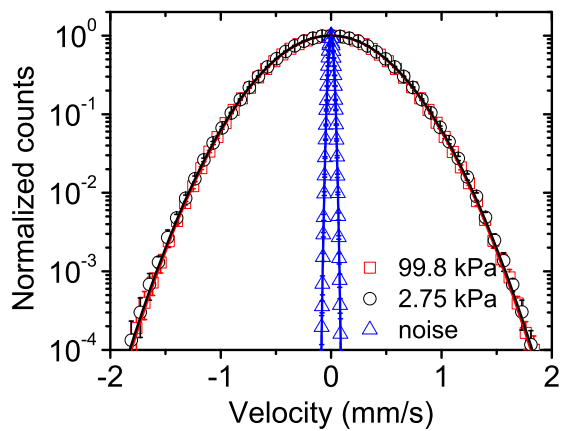}}
\end{picture}
\caption[Measured velocity distribution of a bead]{\label{measuredvelocity}  The distribution of the measured instantaneous velocities of a 3 $\mu$m silica bead. The statistics at each pressure are calculated from 4 million instantaneous velocities. The solid lines are Maxwell-Boltzmann distributions.  Figure adapted from Ref. \cite{li2010}.}
\end{figure}

Figure \ref{measuredtrajectory} shows typical samples of position and velocity traces of a trapped bead.   The position traces of the bead at these two pressures appear very similar.  The instantaneous velocity of the bead at 99.8 kPa changes more frequently than that at 2.75 kPa, because the momentum relaxation time is  shorter at higher pressure.

Figure \ref{MSDinair} shows the mean square displacements of a 3 $\mu$m silica bead as a function of time. The measured MSD's fit  excellently with Eq. \ref{eq4-17} over three decades of time for both pressures.  The measured MSD's are completely different from those predicted by Einstein's theory of Brownian motion in a diffusive regime. The slopes of measured MSD curves at short time scales are double of those of the MSD curves of diffusive Brownian motion in the log-log plot (Fig. \ref{MSDinair}). This is because the  MSD is proportional to $t^2$ for ballistic Brownian motion, and it is proportional to $t$ for diffusive Brownian motion. Another important feature is that the MSD curves are independent of  air pressure at short time scales, as is predicted by Eq. \ref{eq4-14} for ballistic Brownian motion, whereas the MSD in the diffusive regime does depend on the air pressure. At long time scales, the MSD oscillates and saturates at a constant value because of the optical trap.

The distributions of the measured instantaneous velocities are displayed in Fig. \ref{measuredvelocity}. They agree very well with the Maxwell-Boltzmann distribution. The measured rms velocities are $v_{rms}$ = 0.422 mm/s at 99.8 kPa and $v_{rms}$ = 0.425 mm/s at 2.75 kPa. These are very close to the prediction of the energy equipartition theorem, $v_{rms}=\sqrt{k_B T/M}$, which is 0.429 mm/s. As expected, the velocity distribution is independent of pressure.  Thus the Maxwell-Boltzmann distribution of velocities and the equipartition theorem of energy for  Brownian motion were verified directly.

\begin{figure}[t!]
\setlength{\unitlength}{1cm}
\begin{picture}(8,6)
\put(0,0){\includegraphics[totalheight=6cm]{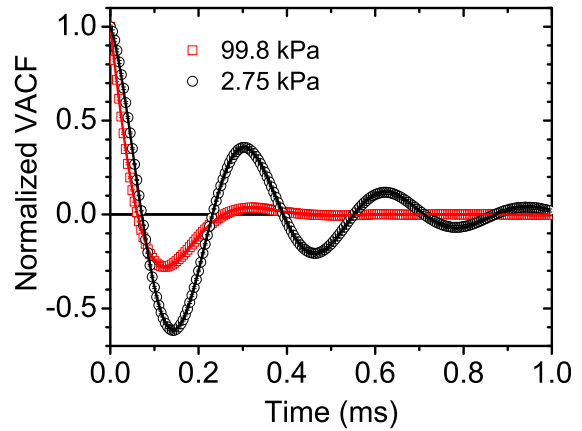}}
\end{picture}
\caption[Measured velocity correlation of a bead]{\label{fig-4-12}  The normalized velocity autocorrelation functions of the 3 $\mu$m bead at 99.8 kPa (red square) and at 2.75 kPa (black circle) from the measurements. The solid lines are fittings with Eq. \ref{eq4-20}. Figure adapted from Ref. \cite{li2010}.}
\end{figure}

\begin{figure}[tp!]
\setlength{\unitlength}{1cm}
\begin{picture}(8,6.5)
\put(0,0){\includegraphics[totalheight=6.5cm]{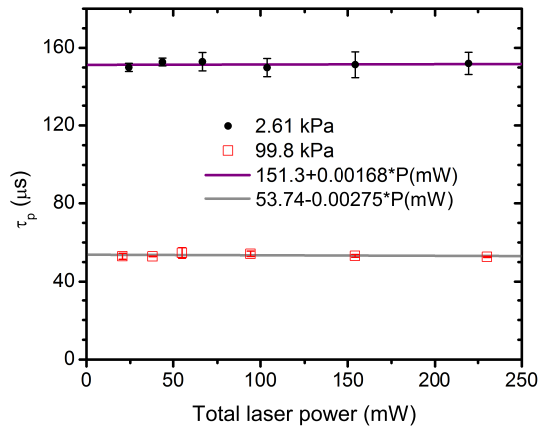}}
\end{picture}
\caption[Measured $\tau_p$ v.s. laser power]{\label{fig-4-13}  Measured momentum relaxation times ($\tau_p$) of a microsphere trapped at 2.61 kPa and 99.8 kPa  as a function of the total  power of the two trapping  beams.}
\end{figure}

Figure \ref{fig-4-12} shows the normalized VACF of the bead at two different pressures. They fit with Eq. \ref{eq4-20} nicely. At 2.75~kPa, one can clearly see the oscillations  due to the optical trap. Eq. \ref{eq4-20} is independent of the calibration factor  of the detection system. The only independent variable is time $t$, which we can measure with high precision. Thus the normalized VACF provides an accurate method to measure  $\tau_p$ and $\omega_0$. We can also calculate the diameter of the silica bead from the $\tau_p$ value at 99.8~kPa \cite{moshfegn2009}. The obtained diameter for this microsphere is 2.79~$\mu$m. This is within the uncertainty range given by the supplier of 3.0~$\mu$m silica beads. We use this value in the calculation of MSD and normalized VACF.

For a particle at a certain pressure and temperature,  $\tau_p$ should be independent of the trapping frequency. We verified this by changing the total power of the two laser beams from 25 mW to 220mW. The measured $\tau_p$ of a microsphere trapped at 19.6 torr and 749 torr  as a function of the total laser power is shown in Fig. \ref{fig-4-13}. Here we use a new microsphere and a smaller data set to calculate the $\tau_p$ than those used in the previous figures. Although the data points for each pressure are not perfectly on a line, it is clear that the  $\tau_p$'s are independent of the laser power within the experimental uncertainty. Fitting the data for each pressure with a straight line, we obtain $\tau_p = [151.3 + 0.00168 \, P/(1 \text{mW})]\, \mu \text{s}$ at 2.61 kPa, and $\tau_p = [53.74 - 0.00275 \, P /(1 \text{mW})]\, \mu \text{s} $ at 99.8 kPa for this microsphere, where $P$ is the total power of the two trapping beams.
Thus $\tau_p$ changes less than 1.3\% for both pressures when the total laser power is changed from 0 to 230 mW. This proves that the fitting method is very accurate, and the heating due to the laser beams (which would change the viscosity \cite{peterman2003,guzman2008} and affect $\tau_p$) is negligible at these pressures.

\section{Brownian motion in a liquid}

The main difference between the Brownain motion in a liquid and that in a gas is the hydrodynamic effects of the liquid \cite{selmeczi2007}.  The Brownian motion of colloidal particles in a liquid at high concentrations have been studied with diffusing wave spectroscopy, which requires each photon to be scattered many times before reaching the detector \cite{weitz1989,zhu1992,kao1993}.  The recent developments in optical tweezers provide a new tool for studying the Brownian motion of single particles with unprecedented precision \cite{lukic2005,chavez2008,li2010,huang2011,franosch2011}. Meanwhile,  precise calibrations (force, position, etc.) of optical tweezers demand better  understanding of the Brownian motion of  trapped particles \cite{bergsorensen2004,schaffer2007,guzman2008,grimm2012}.

\subsection{Theory}
Besides the inertia of the particle itself,  the inertia of the surrounding liquid is also important for  Brownian motion of particles in a liquid. The motion of a particle will cause long-lived vortices in the liquid that will affect the  motion  of the particle itself. This is the hydrodynamic memory effect of the liquid, which dominates the dynamics of the particle at short time scales. These hydrodynamic memory effects were first studied by Vladimirsky in 1945 \cite{vladimirsky1945}. In 1960s, several authors found in computer simulations that the velocity autocorrelation function (VACF) of fluid molecules had a power-law tail in the form of $t^{-3/2}$ \cite{alder1967,alder1970}, in contrast to the exponential decay in a dilute gas. Hinch obtained an analytical solution of the VACF for free particles from the original Langevin analysis \cite{hinch1975}. Clercx and Schram calculated the MSD and VACF of a Brownian particle in  a harmonic potential in an incompressible liquid \cite{clercx1992}, which can be used to describe the Brownian motion of an optically trapped microsphere in a liquid directly \cite{jeney2008,huang2011,franosch2009}.

\subsubsection{A free particle in a liquid}

The effective mass of the microsphere in an incompressible liquid is the sum of the mass of the microsphere and
half of the mass of the displaced liquid \cite{zwanzig1975,brennen1982}:
\begin{equation}
M^* = M_p +  \frac{1}{2} M_f ,
\end{equation}
where $M_p = (4/3)\pi R^3 \rho_{p}$ is the mass of the microsphere, $M_f = (4/3)\pi R^3 \rho_{f}$ is the mass of displaced liquid, $\rho_{p}$ is the density of the microsphere, and $\rho_{f}$ is the density of liquid. The energy equipartition theorem needs to be modified to:
\begin{equation}
\frac{1}{2}M^* \langle v^2 \rangle = \frac{1}{2} k_B T
\end{equation}
where $v$ is the velocity of the microsphere in one dimension. Thus the rms velocity is $v_{rms} = \sqrt{k_B T /M^*}$.
Because of the memory effect of liquid, the velocity autocorrelation function (VACF) of a free particle in a liquid will not be $\langle v(t)v(0)\rangle = \frac{k_B T}{M} e^{-  t/\tau_p}$ as in air, but
 \cite{hinch1975,selmeczi2007,clercx1992,huang2008}
 \begin{equation}
 \label{eq-5-VACF-free-in-water}
\frac{\langle v(t)v(0)\rangle}{k_B T /M^*} =  \frac{\alpha_+ e^{\alpha^2_+ t} \text{erfc}(\alpha_+\sqrt{t})-\alpha_- e^{\alpha^2_- t} \text{erfc}(\alpha_-\sqrt{t})}{\alpha_+ - \alpha_-} ,
\end{equation}
where
\begin{equation}
\alpha_{\pm}= \frac{3}{2} \cdot \frac{3 \pm (5-36 \tau_p/ \tau_f)^{1/2}}{\tau_f^{1/2}(1+9 \tau_p/\tau_f)}.
\end{equation}
$\tau_p=M_p/(6\pi \eta R) = \frac{2}{9} R^2 \rho_{p}/\eta$ is
the momentum relaxation time of the particle due to its own inertia, $\tau_f=  R^2 \rho_{f}/\eta$ characterizes the effect of liquid.  Here $\eta$ is the viscosity of liquid and $R$ is the radius of the microsphere.

 At long time scales, Eq. \ref{eq-5-VACF-free-in-water} approaches
\begin{equation}
\frac{\langle v(t)v(0)\rangle}{k_B T /M^*} \, \propto \frac{1}{t^{3/2}} \; \; \text{for} \; \; t \rightarrow \infty .
\end{equation}
 At short time scales, Eq. \ref{eq-5-VACF-free-in-water} approaches
\begin{equation}
\label{eq-5-VACF-free-in-water-short-limit}
\frac{\langle v(t)v(0)\rangle}{k_B T /M^*} \, = \exp \left( - b \sqrt{t/ \tau_f}\, \right) \; \; \text{for} \; \; t \rightarrow 0 ,
\end{equation}
where $$b = \frac{18}{\sqrt{\pi}(1+2 \rho_p /\rho_f)} \, .$$
For a silica microsphere in water, $b=2.03$. The normalized VACF approaches 1 at
short time scales as \\
$\exp(-b \sqrt{t/\tau_f})$, rather than $\exp(-t /\tau_p)$. Thus the  dynamics of the
particle is dominated by the hydrodynamic effects of the liquid. This is very different from the case in air.

\subsubsection{An optically trapped microsphere in a liquid}

\begin{table}[t!]
	\centering
		\begin{tabular}{|c|c|c|c|c|c|}
        \hline
Diameter & $\tau_p$ & $\tau_f$  & $\tau_c$ &  $k$  & $\tau_k$  \\
\hline
($\mu$m) &($\mu$s)  &  ($\mu$s) & (ns)   &  ($\mu$N/m)  &($\mu$s)     \\
                                \hline
1.0&  0.11        &0.25 &  0.34  & 100 & 94   \\
\hline
3.0  &  1.0        &2.2  &  1.01 & 33.3 & 851  \\
\hline
4.7  &  2.45        &5.51  &  1.58  & 21.3& 2083  \\
\hline
10   &  11.1       &25.0  &  3.4  & 10 & 9443  \\
\hline
	\end{tabular} 
\caption{\label{time-scales-in-water} Characteristic time scales of an optically trapped silica microsphere in water  at 20 $^{\circ}$C.
Some examples of the spring constant of the optical trap ($k$) are shown in the 5th column. It is assumed to be inversely proportional to the diameter of the microsphere when the laser power is constant. }
\end{table}

The optical trap provides a harmonic force $F_{trap} = - k x$ on the microsphere when the displacement of the microsphere is small. $k= M_p \Omega^2$ where $\Omega$ is the natural angular frequency of the trap. Clercx and Schram \cite{clercx1992} gave analytical solutions for the MSD and  VACF of a trapped Brownian particle in a liquid, and Berg-S{\o}rensen and Flyvbjerg \cite{bergsorensen2004} gave a solution for the power spectrum density (PSD) of a trapped Brownian particle  in a liquid. This section introduces their analytical solutions and provides some numerical results to visualize those solutions.

Because the velocity of the Brownian motion of a microsphere in liquid is much smaller than the speed of sound in the liquid,  the fluid motion can be described by the linearized incompressible time-dependent Navier-Stokes equation.
The Langevin equation of the motion of a trapped microsphere in an incompressible liquid  is \cite{clercx1992}:
\begin{eqnarray}
\label{eq-5-navier-stokes}
M^* \ddot{x}(t)&=& -k x(t) - 6 \pi \eta R \dot{x}(t) \\ \nonumber
- 6&R^2&\sqrt{\pi \rho_f \eta} \int_{-\infty}^t (t-t')^{-1/2} \ddot{x}(t') d t' +F_{therm}(t).
\end{eqnarray}
The first term after the equal sign of Eq. \ref{eq-5-navier-stokes} is the harmonic force, the second term is the ordinary Stokes's friction, the third term is a memory term associated with the hydrodynamic retardation effects of the liquid, and the last term is the Brownian stochastic force.

By the famous fluctuation-dissipation theorem, the thermal force is directly related to the frictional force. So the hydrodynamic memory of the liquid will affect both the thermal force and the frictional force. The thermal force is not a white noise, but becomes colored. The correlation in the thermal force is \cite{franosch2011,selmeczi2007,clercx1992,bergsorensen2005}
 \begin{eqnarray}
\label{eq-thermalforce}
\langle F_{therm}(t) F_{therm}(0) \rangle = -\gamma k_B T\sqrt{\frac{\tau_f}{4 \pi}} t^{-3/2},
\end{eqnarray}
which is very different from a delta function (Eq. (\ref{eq4-4})).

\begin{figure}[b!]
\setlength{\unitlength}{1cm}
\begin{picture}(8,6.5)
\put(0.0,0){\includegraphics[totalheight=6.5cm]{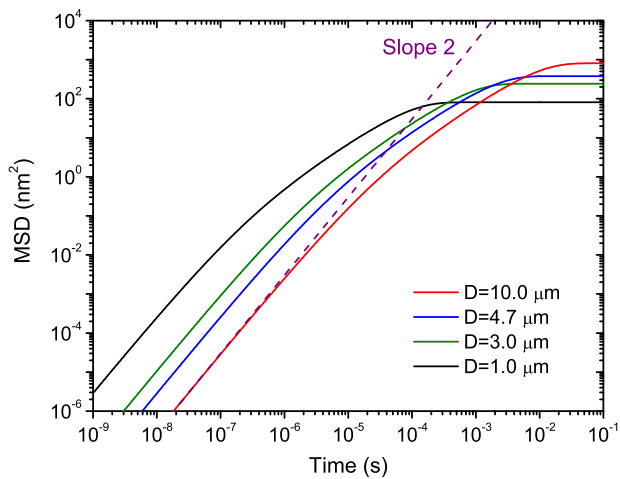}}
\end{picture}
\caption[MSD of a Brownian particle in water]{\label{fig-5-MSDinwater}  Calculated mean square displacement of an optically trapped silica microsphere in water at 20 $^{\circ}$C. Parameters are the same as those in Table \ref{time-scales-in-water}. }
\end{figure}

The mean-square displacement of a trapped microsphere in a liquid is \cite{clercx1992,lukic2007}
\begin{eqnarray}
\label{eq-5-MSD-trapped-particle-in-waater}
\langle [\Delta x(t)]^2 \rangle_{trap} &=& \frac{2 k_B T}{k}  \\ \nonumber
& & + \frac{2 k_B T}{M^*} [
\frac{e^{z^2_1 t} \,\text{erfc}(z_1 \sqrt{t})}{z_1 (z_1-z_2)(z_1-z_3)(z_1-z_4)} \\ \nonumber
& & + \frac{e^{z^2_2 t}\, \text{erfc}(z_2 \sqrt{t})}{z_2 (z_2-z_1)(z_2-z_3)(z_2-z_4)} \\ \nonumber
& & + \frac{e^{z^2_3 t}\, \text{erfc}(z_3 \sqrt{t})}{z_3 (z_3-z_1)(z_3-z_2)(z_3-z_4)} \\ \nonumber
& & + \frac{e^{z^2_4 t}\, \text{erfc}(z_4 \sqrt{t})}{z_4 (z_4-z_1)(z_4-z_2)(z_4-z_3)}
]
\end{eqnarray}
The coefficients $z_1$, $z_2$, $z_3$, and $z_4$ are the four roots of the equation \cite{lukic2007}
\begin{equation}
\left(\tau_p+\frac{1}{9}\tau_f \right)z^4-\sqrt{\tau_f}z^3+z^2+\frac{1}{\tau_k} =0 ,
\end{equation}
where $\tau_k = 6 \pi \eta R / k$. For $t \rightarrow \infty$, Eq. \ref{eq-5-MSD-trapped-particle-in-waater} approaches $$\langle{[\Delta x(\infty)]^2}\rangle_{trap} = \frac{2 k_B T}{k} .$$

\begin{figure}[b!]
\setlength{\unitlength}{1cm}
\begin{picture}(8,6.5)
\put(0.0,0){\includegraphics[totalheight=6.5cm]{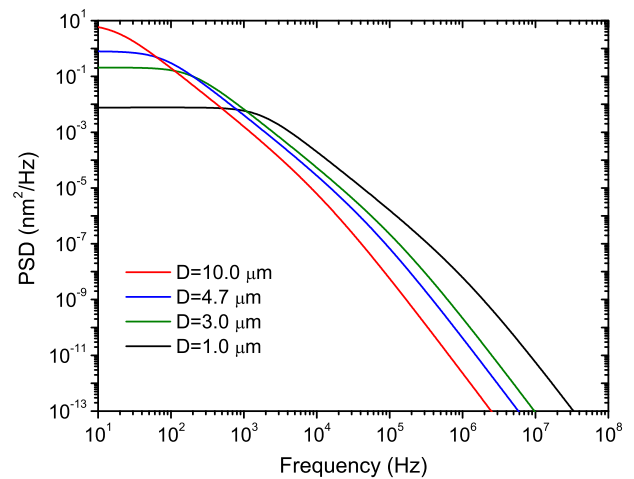}}
\end{picture}
\caption[PSD of a Brownian particle in water]{\label{fig-5-PSDinwater}  Calculated power spectra of an optically trapped silica microsphere in water at 20 $^{\circ}$C. Parameters are the same as those in Table \ref{time-scales-in-water}. }
\end{figure}

The normalized VACF of a trapped microsphere in a liquid is \cite{clercx1992,lukic2007}
 \begin{eqnarray}
\label{eq-5-VACF-trapped-particle-in-water}
A(t) = \frac{\langle v(t)v(0) \rangle}{k_B T/M^*}&=&
\frac{z^3_1 \, e^{z^2_1 t} \, \text{erfc}(z_1 \sqrt{t})}{(z_1-z_2)(z_1-z_3)(z_1-z_4)} \\ \nonumber
& & + \frac{z^3_2 \,e^{z^2_2 t} \,\text{erfc}(z_2 \sqrt{t})}{ (z_2-z_1)(z_2-z_3)(z_2-z_4)} \\ \nonumber
& & + \frac{z^3_3 \,e^{z^2_3 t} \,\text{erfc}(z_3 \sqrt{t})}{ (z_3-z_1)(z_3-z_2)(z_3-z_4)} \\ \nonumber
& & + \frac{z^3_4 \,e^{z^2_4 t} \,\text{erfc}(z_4 \sqrt{t})}{ (z_4-z_1)(z_4-z_2)(z_4-z_3)} .
\end{eqnarray}

\begin{figure}[bp!]
\setlength{\unitlength}{1cm}
\begin{picture}(8,6.5)
\put(0,0){\includegraphics[totalheight=6.5cm]{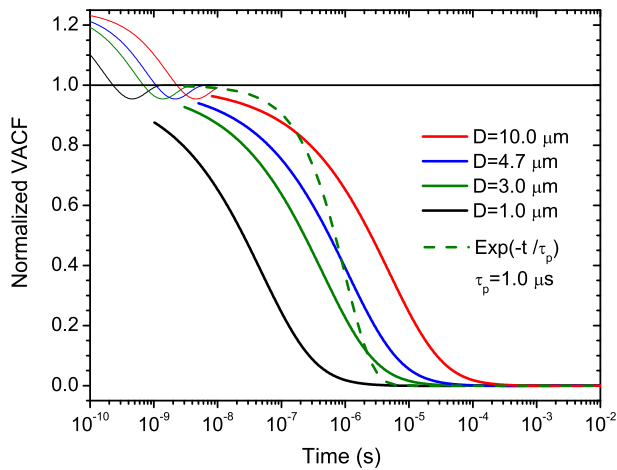}}
\end{picture}
\caption[VACF of a Brownian particle in water]{\label{fig-5-VACFinwater}  Calculated normalized velocity autocorrelation function of an optically trapped silica microsphere  in water at 20 $^{\circ}$C. Parameters are the same as those in Table \ref{time-scales-in-water}.  The thin solid lines ($t < 10^{-8}$~s) are calculated from Eq. \ref{eq-5-VACF-short-in-water}, and the thick solid lines are calculated from Eq. \ref{eq-5-VACF-trapped-particle-in-water}. The dashed lines are exponential decays with $\tau_p=1.0 \mu$s, corresponding to a 3.0 $\mu$m microsphere. }
\end{figure}

The power spectral density is \cite{bergsorensen2004,lukic2007}:
\begin{eqnarray}
\label{eq-5-PSD-water}
&&S(f) = \frac{D}{2\pi^2 f^2} \cdot \\ \nonumber
&&\frac{1+\sqrt{f/2 \phi_f}}{(\phi_k/f-\sqrt{f/2\phi_f}-f/\phi_p-f/9\phi_f\,)^2+(1+\sqrt{f/2\phi_f}\,)^2} \, ,
\end{eqnarray}
where $f$ is the observation frequency, $\phi_{k}=1/(2\pi \tau_{k})$ is the corner frequency of the power spectrum due to the trap, and $\phi_{p,f}=1/(2\pi \tau_{p,f})$.
For $f \rightarrow 0$, Eq. \ref{eq-5-PSD-water} approaches
$$S(0) = \frac{2 k_B T \gamma}{k^2},$$
where $\gamma= 6 \pi \eta R$.

At $t \rightarrow 0$, Eq. \ref{eq-5-VACF-trapped-particle-in-water} predicts $\langle v(0)v(0) \rangle$ = $k_B T/M^*$,
which is different from the energy equipartition theorem
$\langle v(0)v(0) \rangle$ = $k_B T/M_p$. This conflict is caused by the assumption in Eq. \ref{eq-5-navier-stokes} that the liquid is incompressible. For $t < t_c$, we need to consider the liquid to be compressible. Here $t_c = R /c$ is the time required for a sound wave to travel a sphere radius, where $c$ is the speed of sound in the liquid. The effects of compressibility have been studied by Zwanzig and  Bixon \cite{zwanzig1975}. The normalized velocity autocorrelation function at $t \sim t_c$ is \cite{zwanzig1975}:
 \begin{eqnarray}
\label{eq-5-VACF-short-in-water}
A(t) &=& \frac{\langle v(t)v(0) \rangle}{k_B T/M^*}\\ \nonumber
&=& 1 + \frac{M_f}{2 M_p}\left[ \frac{1}{2}-\frac{i M^*}{(4 M_p^2-M_f^2)^{1/2}}\right] e^{-i x_1 t/t_c} \\ \nonumber
& & + \frac{M_f}{2 M_p}\left[ \frac{1}{2}+\frac{i M^*}{(4 M_p^2-M_f^2)^{1/2}}\right] e^{-i x_2 t/t_c} ,
\end{eqnarray}
where
 \begin{eqnarray}
x_1 =- i \frac{M^*}{M_p} + [1-\frac{M_f^2}{4 M_p^2}]^{1/2}, \\
x_2 =- i \frac{M^*}{M_p} - [1-\frac{M_f^2}{4 M_p^2}]^{1/2}.
\end{eqnarray}
At very short time scales $t \ll t_c$, Eq. (\ref{eq-5-VACF-short-in-water}) approaches $A(0)=1+ \frac{M_f}{2 M_p}$. The short time limit $A(0)\neq 1$ because the normalization factor is $k_B T/M^*$ in Eq. (\ref{eq-5-VACF-short-in-water}), rather than  $k_B T/M_p$.

The calculated MSD's of microspheres with different diameters in water are shown in Fig. \ref{fig-5-MSDinwater}, and the corresponding power spectra are shown in Fig. \ref{fig-5-PSDinwater}. Fig. \ref{fig-5-VACFinwater} displays the normalized velocity autocorrelation function ($A(t)$) of an optically trapped silica microsphere  in water at 20 $^{\circ}$C. The thick solid lines are calculated from Eq. \ref{eq-5-VACF-trapped-particle-in-water}, which treats the water as an incompressible fluid. The thin solid lines at short time scales ($t < 10^{-8}$~s) are calculated from Eq. \ref{eq-5-VACF-short-in-water}, which includes the compressibility effects of water.
The dashed lines are exponential decays with $\tau_p=1.0 \, \mu$s, corresponding to a 3.0~$\mu$m microsphere. As clearly shown in Fig. \ref{fig-5-VACFinwater}, the VACF of a microsphere in water is very different from exponential decay because of the hydrodynamic memory effects of water.

The thin solid lines are expected to be correct for $t \sim t_c$, and the thick sold lines are expected to be correct for $t \gg t_c$. The intermediate regime $t_c < t < 100 t_c$ is poorly understood.
A recent experiment has measured the VACF of a Brownian particle in water at VACF$ < 0.35$ \cite{huang2011,jeney2008}. A measurement of the VACF between 1 and 0.35 is required in order to better understand the hydrodynamic effects and compressibility effects of water on  Brownian motion \cite{metiu1977,felderhof2007,schram1998,erbas2010}.

\subsection{Experimental observation of the color of thermal force in a liquid}

\begin{figure}[b!]
\setlength{\unitlength}{1cm}
\begin{picture}(8,12)
\put(0.8,0){\includegraphics[totalheight=12cm]{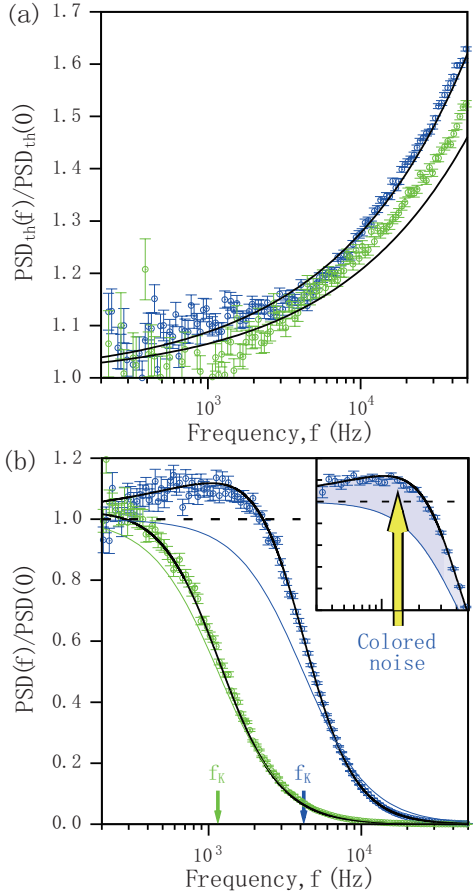}}
\end{picture}
\caption[The color of thermal force]{\label{thermalforce}  The color of thermal force (Figure adapted from Ref. \cite{franosch2011}). (a), The normalized power spectral density (PSD) of thermal force of an optically trapped melamine resin sphere (R = 1.45 $\mu$m) in water (green circles) or acetone (blue circles).  The black lines are predictions of the hydrodynamic theory. (b), The normalized power spectral density of position.  }
\end{figure}

Conventionally, the thermal force exerted on a Brownian particle  are assumed to be a white noise.
Due to the hydrodynamic memory effects of the liquid, this thermal force is in fact colored. Direct experimental observation of the color of the thermal force in a liquid was elusive until the recent work by Franosch \emph{et al} \cite{franosch2011}. They used optical tweezers with very large stiffness ($k \approx 205\, \mu$N/m for 2.9 $\mu$m diameter particles) to study the Brownian motion of a trapped microsphere. Their key insight is that at long times, strong trapping eventually dominates over friction and becomes the main force counteracting thermal force \cite{bergsorensen2005}. So the Langevin equation reduces to $kx(t)\approx F_{therm}(t)$. Consequently, the correlations in the thermal force can be obtained by the position autocorrelation function $\langle F_{therm}(t)F_{therm}(0)\rangle \approx k^2 \langle x(t)x(0)\rangle$.

Figure \ref{thermalforce} shows the  results of Franosch \emph{et al}'s experiment on the color of thermal force \cite{franosch2011}. The data in Fig. \ref{thermalforce}(a) clearly shows the departure of thermal force from a white noise whose power spectrum is a horizontal line. The measured spectrum of the thermal force increases at higher frequencies. Franosch \emph{et al} also observed resonances in Brownian motion in liquid where overdamped motion is often assumed. In order to observe the resonance, they used optical tweezers with very large stiffness, and acetone as a liquid rather than water. The viscosity of acetone is  about 3 times smaller than that of water. So the motion of a trapped bead is less damped in acetone than in water. The resonance can be clearly seen in Fig. \ref{thermalforce}(b). Remarkably, this resonance is mainly due to the inertia of the liquid, rather than the inertia of the particle itself.

\subsection{Experimental observation of the transition from ballistic to diffusive Brownian motion in a liquid}

\begin{figure}[b!]
\setlength{\unitlength}{1cm}
\begin{picture}(8,6)
\put(0,0){\includegraphics[totalheight=6cm]{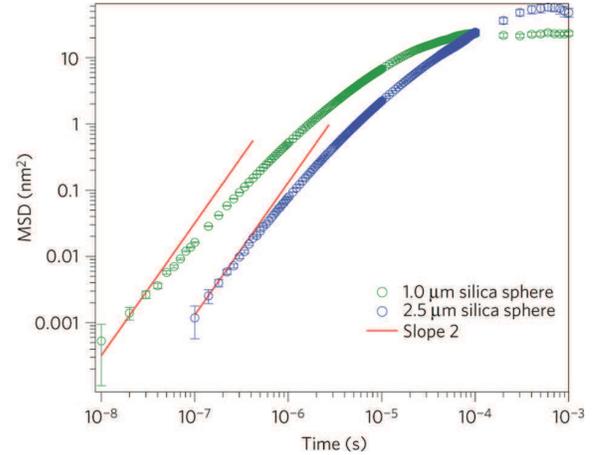}}
\end{picture}
\caption[Measured mean square displacement of a bead]{\label{MeasuredMSDinwater}  Measured mean square displacement of silica spheres with diameters of 1 $\mu$m and 2.5 $\mu$m (Figure adapted from Ref. \cite{huang2011}).  The red lines with slope 2 show the  expected behavior of ballistic Brownian motion. }
\end{figure}

Recently, Huang \emph{et al} studied the Brownian motion of a single particle in an optical trap in water with sub-Angstrom resolution and measured the velocity autocorrelation function of the Brownian motion \cite{huang2011}.   In their experiments, $\tau_k$ due to the confinement of the optical tweezers was typically two orders of magnitude larger than $\tau_f$. So the role of the optical confinement can be neglected during the transition from ballistic to diffusive Brownian motion.

\begin{figure}[b!]
\setlength{\unitlength}{1cm}
\begin{picture}(8,6)
\put(0,0){\includegraphics[totalheight=6cm]{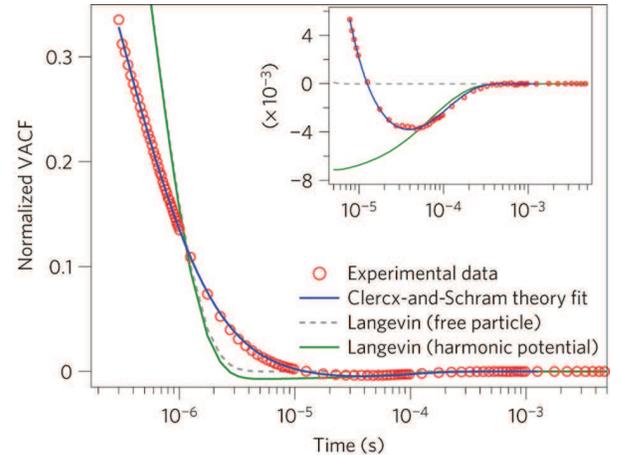}}
\end{picture}
\caption[VACF]{\label{measuredVACF}  Normalized experimental VACF for a 2 $\mu$m diameter resin sphere in an optical trap (Figure adapted from Ref. \cite{huang2011}).  The blue line shows the fitting of the Clercx-and-Schram theory (Eq. (\ref{eq-5-VACF-trapped-particle-in-water})) to the whole experimental VACF.  The grey (dashed) and green lines show the expectations of VACF for the same particle if the inertia of the liquid is ignored,   in the absence (grey) and presence (green) of the harmonic trap. }
\end{figure}

Figure \ref{MeasuredMSDinwater} shows Huang \emph{et al}'s results of MSD's for  1 $\mu$m and  2.5 $\mu$m silica particles from 10 ns to 1 ms \cite{huang2011}.  The MSD's increase at short time scales, and reach plateaus at long time scales due to  the confinement of the optical trap. At short time scales, the particle undergoes free, but correlated Brownian motion because of the inertia of the particle and the surrounding liquid.  The effect of the particles' inertia on Brownian motion is visible in Fig. \ref{MeasuredMSDinwater}. The slopes of the MSDs are close to 2 at short time scales, as predicted for ballistic Brownian motion. It is remarkable that Huang \emph{et al} achieved a temporal resolution of  about 10 ns in the case of a 1 $\mu$m silica particle. At this temporal resolution, Huang \emph{et al} resolved a MSD as small as ~0.0005 nm$^2$, corresponding to ~20 pm spatial resolution. The temporal resolution shown in Fig. \ref{MeasuredMSDinwater} is about 0.1 $\tau_p$ for a 1 $\mu$m particle, and about 0.15 $\tau_p$ for a 2.5 $\mu$m particle.

Figure \ref{measuredVACF} shows Huang \emph{et al}'s experimental results of the VACF of a 2 $\mu$m resin particle and a theoretical fit with the Clercx-and-Schram theory (Eq. (\ref{eq-5-VACF-trapped-particle-in-water})) \cite{clercx1992}. The predictions of VACF from Langevin equations neglecting the hydrodynamic effects of water are also displayed for comparison.  The experimental results agree with the Clercx-and-Schram theory. For times shorter than $\tau_f = 1 \,\mu$s, the velocity correlations are smaller than the predictions of the Langevin equation neglecting  the inertia of the fluid. At longer times, the correlations are stronger because of the vortex developed in the fluid. The small anti-correlation dip near 300 $\mu$s is because of the optical trap.

 Although the spatial resolution of Huang \emph{et al}'s experiment is sufficient to observe the transition from ballistic to diffusive Brownian motion in MSD and even compute a VACF, it is not enough to measure the instantaneous velocity of  Brownian motion in a liquid. The reason is that both MSD and VACF are insensitive to white noise of the detection system since they are averages over large data sets. On the other hand, the measurement of the instantaneous velocity is very sensitive to the  detection noise. We will discuss  about this issue in the following section.

\section{Effects of detection noise on studying  Brownian motion}

In the presence of  detection noise \cite{chavez2008,li2010,huang2011}, the measured position of the microsphere can be expressed as
\begin{equation}
x_{msr}(t)= x_p(t) + x_n(t),
\end{equation}
where $x_p(t)$ is the real position of the microsphere, and $x_n(t)$ is the noise of the detection system. The mean square displacement (MSD) of the measured positions is \cite{huang2008}:
\begin{eqnarray}
&&MSD_{msr}(t) = \langle [x_{msr}(t_0 + t)-x_{msr}(t_0 )]^2  \rangle \\ \nonumber
 &&= \langle [x_{p}(t_0 + t)-x_{p}(t_0 )]^2  \rangle  + \langle [x_{n}(t_0 + t)-x_{n}(t_0 )]^2 \rangle \\  \nonumber
 &&   + 2 \langle [x_{p}(t_0 + t)-x_{p}(t_0 )] \cdot [x_{n}(t_0 + t)-x_{n}(t_0 )] \rangle \\ \nonumber
 &&= MSD_p(t) + MSD_n(t),
\end{eqnarray}
where the average is taken over all possible $t_0$. This derivation assumes no correlation between the real position of the microsphere and the detection noise. In this case, the real MSD of the microsphere can be obtained by $MSD_p(t)= MSD_{msr}(t)-MSD_n(t)$, as is done by Huang \emph{et al} in Ref. \cite{lukic2005,huang2011}. $MSD_n(t)$ is usually independent of time, as shown in Fig. \ref{MSDinair}.

 The measured velocity of the microsphere is
\begin{eqnarray}
&&v_{msr}(t) =  \frac{x_{msr}(t+\frac{\Delta t}{2})-x_{msr}(t-\frac{\Delta t}{2})}{\Delta t} \\ \nonumber
&&=  \frac{x_{p}(t+\frac{\Delta t}{2})-x_{p}(t-\frac{\Delta t}{2})}{\Delta t} +
 \frac{x_{n}(t+\frac{\Delta t}{2})-x_{n}(t-\frac{\Delta t}{2})}{\Delta t}\\ \nonumber
&&= v_p(t) + v_n(t),
\end{eqnarray}
where $\Delta t \ll \tau_p$. Because the measured velocity contains a noise signal $v_n(t)$, the smallest  $\Delta t$ of the detection system may not be the best value  for measuring the velocity. The data acquisition (DAQ) card creates noise when it converts an analog signal to a digital signal due to the finite number of bits. The minimum value of $x_{n}(t+\frac{\Delta t}{2})-x_{n}(t-\frac{\Delta t}{2})$ is limited by the DAQ card, thus $v_n(t)$ may be larger than the real velocity of the microsphere ($v_p(t)$) if $\Delta t$ is too small.

 The measured velocity represents the real instantaneous velocity of the microsphere if $\Delta t \ll \tau_p$ and $v_n(t)$ is negligible. This requires $\langle v^2_{msr} \rangle \gg \langle v^2_{n} \rangle$. One can check whether this condition is satisfied by comparing the signal when a microsphere is trapped in the optical tweezer and when there is no microsphere in the optical tweezer.  The relation between $\langle v^2_{msr} \rangle$ and $\langle v^2_{n} \rangle$ can  be obtained from the measured $MSD$'s:
\begin{eqnarray}
\langle v^2_{msr} \rangle &=&  \langle \frac{ [x_{msr}(t+\frac{\Delta t}{2})-x_{msr}(t-\frac{\Delta t}{2})]^2 }{\Delta t^2} \rangle \\ \nonumber
&=& \frac{MSD_p(\Delta t)}{\Delta t ^2} + \frac{MSD_n(\Delta t)}{\Delta t ^2} \\ \nonumber
&=& \langle v^2_{p} \rangle +\langle v^2_{n} \rangle.
\end{eqnarray}
Thus $\langle v^2_{msr} \rangle \gg \langle v^2_{n} \rangle$ is equivalent to
$MSD_{msr}(\Delta t) \gg MSD_n(\Delta t)$.

The measured velocity autocorrelation function is
\begin{eqnarray}
\langle v_{msr}(t+t_0) v_{msr}(t_0) \rangle &=& \langle v_p(t+t_0) v_p(t_0) \rangle \\ \nonumber
&+&\langle v_n(t+t_0) v_n(t_0) \rangle .
\end{eqnarray}
If the  noise of the detection system  has almost no correlation (white noise), the last term of this equation can be neglected. Thus
\begin{eqnarray}
\langle v_{msr}(t+t_0) v_{msr}(t_0) \rangle =
\langle v_p(t+t_0) v_p(t_0) \rangle .
\end{eqnarray}
So the measurement of the velocity autocorrelation function is not sensitive to the noise of the detection system \cite{huang2011}. On the other hand, the measurement of the instantaneous velocity is very sensitive to the noise of the detection system \cite{li2010}.

If the detection system samples the position of the microsphere every $dt$  that is much shorter than the required temporal resolution $\Delta t$, we can reduce the noise in the measured velocity by using successively averaged positions to calculate the velocity. Let $\Delta t= N \, dt$, where $N \ll \tau_p /dt$, then
\begin{equation}
x_{avr}(t) = \frac{1}{N} \sum_{j=1}^N x_{msr}(t+j dt-\frac{N+1}{2} dt).
 \end{equation}
The measured velocity  becomes
 \begin{equation}
v_{msr}(t) =  \frac{x_{avr}(t+\frac{\Delta t}{2})-x_{avr}(t-\frac{\Delta t}{2})}{\Delta t}.
 \end{equation}
Then the velocity noise is
 \begin{eqnarray}
 \label{eq-averagenoise}
v_{n}(t) &=&  \frac{1}{N^2 dt}[\sum_{j=1}^N x_{n}(t+j dt-\frac{N+1}{2} dt+\frac{\Delta t}{2}) \\ \nonumber
& & -\sum_{j=1}^N x_{n}(t+j dt-\frac{N+1}{2} dt-\frac{\Delta t}{2})].
 \end{eqnarray}
 On average, the rms amplitude of $v_{n}(t)$ in Eq. (\ref{eq-averagenoise}) is  $N \sqrt{N}$ times smaller than that of $[x_n(t+dt/2)-x_n(t-dt/2)]/dt$ if  the position noise $x_n(t)$ is a white noise. If $N=10$, the noise can be reduced by a factor of 32, which is a significant number. The condition of using the averaging method is that $dt \ll \tau_p /N$. So the detection system must have a very high temporal resolution. This averaging method has been utilized in both  the measurement of the VACF of a Brownian particle in a liquid \cite{huang2011} and the measurement of the instantaneous velocity of a Brownian particle in air \cite{li2010}.

\section{Future}

As shown in this review, optical tweezers have become an indispensable tool for studying Brownian motion at short time scales.
The instantaneous velocity of a Brownian particle trapped in a gas has been measured \cite{li2010}.
Recently, the velocity autocorrelation function of a Brownian particle in water was measured successfully for $\frac{\langle v(t)v(0)\rangle }{ \langle v^2\rangle} < 0.35$ \cite{jeney2008,huang2011,huang2008}.

The instantaneous velocity of a Brownian particle in water is much more difficult to measure, and has not been measured to date.
  A successful measurement of the instantaneous velocity of a Brownian particle in a liquid will  complete  the task  that was considered by Einstein more than 100 years ago \cite{einstein1907,einstein1956} and open a new door for studying Brownian motion. For example, the results can be used to test the modified energy equipartition theorem $\langle v(0)v(0) \rangle$ = $k_B T/M^*$ and the Maxwell-Boltzmann velocity distribution. A measurement of the VACF  $>$ 0.35 will deepen our understanding of the hydrodynamic effects and compressibility effects of a liquid on  Brownian motion \cite{metiu1977,felderhof2007,schram1998,erbas2010}.
 The ability to measure the instantaneous velocity of a Brownian particle will be invaluable in studying nonequilibrium statistical mechanics \cite{kubo1986,wang2002}. The Brownian motion of a suspended particle can be used for microrheology to probe the properties of  fluids, such as viscoelastic fluids \cite{atakhorrami2005,grimm2011,Indei2012}, and surrounding environments \cite{tischer2001,tseng2002,Swank2012}.

\begin{acknowledgements}
M.G.R. acknowledges support from the Sid W. Richardson
Foundation and the R. A. Welch Foundation grant number F-1258. We would like to thank S. Kheifets and D. Medellin for their help in the original experiment which  is an important part of this review.
\end{acknowledgements}


%
%

\bibliography{librownian}

\end{document}